\documentclass[aps,prl,twocolumn,showpacs,superscriptaddress]{revtex4-1}
\usepackage{graphicx,amssymb,amsfonts,amsmath}

\begin{document}
\newcommand{\AR}[1]{{\bf AR} {\it #1} {\bf END}}
\newcommand{\OLD}[1]{{\bf OLD} {\tiny #1}}
\newcommand{\sgn}{{\rm sign}}

\title{Mode coupling induced dissipative and thermal effects at long times after a quantum quench}
\author{Aditi Mitra}
\affiliation{Department of Physics, New York University, 4 Washington
  Place, New York, New York 10003, USA}
\author{Thierry Giamarchi}
\affiliation{DPMC-MaNEP, University of Geneva, 24 Quai Ernest-Ansermet, CH-1211 Geneva, Switzerland}
\date{\today}


\begin{abstract}
An interaction quench in a Luttinger liquid can drive it into an athermal steady state.
We analyze the effects on such an out of equilibrium state of a mode coupling term due to
a periodic potential. Employing a
perturbative renormalization group approach we show that even when the periodic potential is
an irrelevant perturbation in equilibrium, it has important consequences on the athermal steady state
as it generates a temperature as well as a dissipation and hence a finite life-time
for the bosonic modes.
\end{abstract}


\maketitle

The high degree of tunability and control associated with cold-atomic gases make
them an exciting testbed for studying a host of phenomena related to interacting quantum
particles~\cite{Bloch08}. Among the topics of great current interest are the non-equilibrium
physics of quantum quenches for which these systems are particularly well
adapted (see~\cite{Polkovrev09}
and references therein) and
other classes of steady state non-equilibrium phenomena such as
systems subjected to a time dependent noise~\cite{Torre10}.

Crucial questions in all these out of equilibrium phenomena is the relaxation mechanism by which the
system reaches a steady state and the properties of the steady state, in particular whether
the latter is athermal or thermal and hence described by a Gibbs distribution.
In the case of a quench where non-trivial time-evolution is triggered after a sudden change
in system parameters, many models mainly in one-dimension~\cite{Rigol07}
and involving simple effective theories~\cite{Cazalilla06,Calabrese06}, are found to reach an athermal
steady state characterized by a generalized Gibbs ensemble (GGE).
However the generality of the GGE remains under debate since
certain observables do not obey it~\cite{Cazalilla06,Kollath07,Barthel08,Gangardt08,Rossini10}.
More generally, the relaxation mechanism and the nature of the steady state in more
complicated field-theories, as well as the role of a
finite system size in numerical studies, are still largely unknown~\cite{Roux09,Biroli10}.

It is thus important to have theoretical models for which
such non-equilibrium questions can be reliably studied.
One good candidate for such an analysis is a one dimensional system of interacting bosons,
leading to the so called Luttinger liquid physics~\cite{Giamarchibook}.
The excitations of such a system can be represented by
density modes which are essentially independent. On such a system,
quenches corresponding to a change of the interaction reveal a steady state which
still has independent modes, but these are now characterized by a non-equilibrium distribution that does
not relax to a thermal state~\cite{Cazalilla06}.

In this paper we examine the effect of a mode coupling term on the above system, where the mode coupling is
due to an externally imposed optical lattice.
We address explicitly the question of thermalization
and asymptotic relaxation of such a system. We use the Keldysh technique~\cite{Kamenevrev}
and a controlled renormalization analysis and show that even in cases for which the lattice
potential would be irrelevant in equilibrium, it leads for the out of
equilibrium situation to the appearance of finite
dissipation, as well as a finite temperature for the low energy modes.
We thus explicitly obtain a mechanism, which we argue should be generic, in which
thermalization and dissipation arises due to the transfer of
energy from the long wavelength modes to the short wavelength modes, the latter thus acting as a
bath.

We consider interacting one dimensional bosons in the continuum. The low energy properties of
such a system can
be efficiently represented by a Luttinger liquid~\cite{Giamarchibook}
\begin{equation}\label{Hi}
H_i = \frac{u_0}{2\pi}\int dx
\left[K_0\left(\pi \Pi(x)\right)^2 + \frac{1}{K_0}\left(\partial_x \phi(x)\right)^2
\right]
\end{equation}
where $\phi$ is related to the long wavelength part of the density by $\rho(x)$=$-\nabla\phi(x)/\pi$,
while $\Pi$ is the canonically conjugate variable to $\phi$. The eigenmodes of the Hamiltonian
are the sound waves of density with a dispersion $\omega$=$u_0 q$. The information about
the interactions is contained in the two Luttinger parameters: $u_0$ the velocity of density
oscillations, and $K_0$ a dimensionless parameter controlling the decay of correlation functions.

The bosons are driven out of equilibrium via a sudden interaction quench which for
the effective Luttinger liquid theory, simply implies a sudden change of
the Luttinger parameter from $K_0 \rightarrow K$,
and the velocity from $u_0 \to u$. To satisfy Galilean invariance we choose
$u$=$v_F/K$ and $u_0$=$v_F/K_0$.
The time evolution of the initial state is therefore due to
$H_f$=$H_i(K_0\rightarrow K, u_0 \rightarrow u)$.

We first give here the full solution for a quench $K_0 \rightarrow K$ with arbitrary interactions.
In the Keldysh formalism~\cite{Kamenevrev} it is convenient to define classical
($\phi_{cl}$=$(\phi_-+\phi_+)/\sqrt{2}$)
and quantum ($\phi_q$=$(\phi_--\phi_+)/\sqrt{2}$) fields where $\phi_{-(+)}$ are the time (anti-time)
ordered fields on the Keldysh contour.
In terms of these fields, the action that describes the steady state behavior at long times
($t+t^{\prime}\rightarrow \infty$) after the
quench when transients related to oscillations of $e^{-iu|q|(t+t^{\prime})}$ have averaged out to zero,
is
\begin{multline}\label{S0a}
S_0 = \frac{1}{\pi K u}\sum_{q\neq0,\omega}\begin{pmatrix} \phi_{cl}^* & \phi_q^*\end{pmatrix} \\
\times \begin{pmatrix} 0&& (\omega-i\delta)^2 - u^2 q^2       \\
(\omega + i \delta)^2 - u^2 q^2
&& 4i|\omega|\delta\frac{K_0}{2K}\left(1 + \frac{K^2}{K_0^2}\right)
\end{pmatrix}
\begin{pmatrix}
\phi_{cl}\\
\phi_q
\end{pmatrix}
\end{multline}
Eq.~(\ref{S0a}) implies that the retarded propagator $-i\langle \phi_{cl}\phi_q^*\rangle
= G^R(q,\omega) = \frac{\pi K u}{\left(\omega + i\delta\right)^2 - u^2 q^2}$ is identical 
to that in the ground state of $H_f$, while the Keldysh propagator 
$G^K=-i\langle \phi_{cl}\phi_{cl}^*\rangle$ which is sensitive to the
occupation of the bosonic modes is, 
\begin{eqnarray}
G^K(q,\omega)=
\frac{K_0}{2K}\left(1 + \frac{K^2}{K_0^2}\right) \sgn(\omega)\left[G_R - G_A\right]
\end{eqnarray}
Thus the fluctuation-dissipation theorem (FDT) defined by $G^K$=$(G^R-G^A)\coth(\omega/2T)$, 
where $T$ is the temperature of the bosons, is violated.
When $K$=$K_0$, FDT is recovered as $\coth(\omega/2T)\xrightarrow{T=0} \sgn(\omega)$.
Note that although the system is now out of equilibrium, each $q$ mode is still infinitely
long lived since $\delta$=$0^+$.
\begin{figure}
\includegraphics[totalheight=7cm]{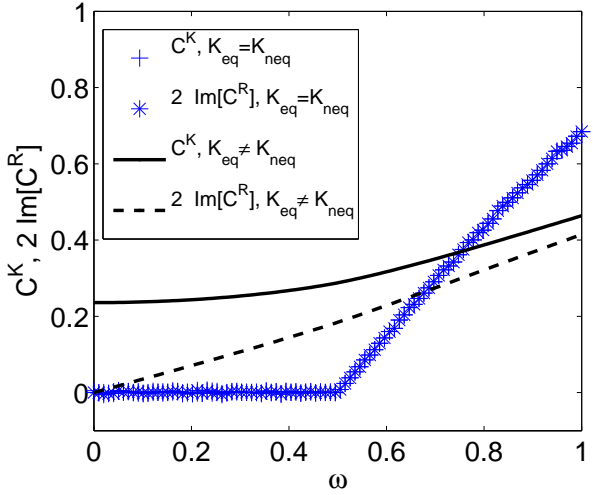}
\caption{\label{fdt} $C^K(q=0.5,\omega)$ and $2{\rm Im}[C^R(q=0.5,\omega)]$
in equilibrium $K_{eq}=K_{neq}=2.0$ and after a quench $K_{eq}=2,K_{neq}=3$}
\end{figure}

$S_0$ can be used to evaluate the basic two-point correlation functions corresponding
to the density fluctuation $C^K_{\phi\phi}$ and response $C^R_{\phi\phi}$, defined as
$C^K_{\phi\phi}(r,t)= -i{\rm Re}\left[e^{-\frac{\gamma^2}{2}\langle (\phi_-(r,t)-\phi_+(0,0))^2\rangle}
\right]$
and
$C^R_{\phi\phi}(r,t)
= i\theta(t){\rm Im}\left[e^{-\frac{\gamma^2}{2}\langle (\phi_-(r,t)-\phi_+(0,0))^2\rangle}\right]$.
Defining $K_{eq}$=$\frac{\gamma^2}{4} K$, $K_{neq}$=$\frac{\gamma^2}{8}K_0(K^2/K_0^2+1)$, we find,
\begin{eqnarray}
&&e^{-\frac{\gamma^2}{2}\langle (\phi_-(r,t)-\phi_+(0,0))^2\rangle} = \nonumber \\
&&e^{-\frac{K_{neq}}{2}\ln\left(\frac{\alpha^2 + \left(ut+r\right)^2}{\alpha^2}\right)-
\frac{K_{neq}}{2}\ln\left(\frac{\alpha^2 + \left(ut-r\right)^2}{\alpha^2}\right)}\nonumber\\
&&e^{i\left[K_{eq}\tan^{-1}\left(\frac{ut+r}{\alpha}\right) + K_{eq} \tan^{-1}
\left(\frac{ut-r}{\alpha}\right)
\right]}\label{corrdef}
\end{eqnarray}
where $\alpha$ is a short distance cut-off, and $\gamma$ is an arbitrary coefficient which
will later be related to the periodicity of an externally imposed lattice potential.
In equilibrium
$K_{eq}$=$K_{neq}$, and one recovers the usual power-law decay of Luttinger liquids.
However out of equilibrium one finds power-law behavior
with new decay exponents $K_{neq}$. For the case of $K_0$=$1$ this
power-law decay was obtained in~\cite{Cazalilla06}, and
can also be recovered using a GGE
that accounts for the conservation of the occupation number of appropriate bosonic modes.
Since $K_{neq}>K_{eq}$, the propagators always decay faster than in equilibrium.

The role played by
the oscillating terms in Eq.~(\ref{corrdef}) which differentiates between response and correlation
functions has not been explored before, and will play an important role in the RG. Its importance
can already be seen at this level by studying the FDT ratio
defined by
$\frac{C^K_{\phi\phi}(q,\omega)}{2{\rm Im}\left[C^R_{\phi\phi}(q,\omega)\right]}$.
While for $K_{eq}$=$K_{neq}$
this ratio reduces to the equilibrium
$T$=$0$
result of ${\rm sign}(\omega)$, out of equilibrium it can be used to
formally define a  $\omega,|q|$ dependent
effective ``temperature''.
In the limit
$\omega\rightarrow 0, q\rightarrow 0$, the effective temperature $\bar{T}$ defined as
\begin{eqnarray}
\frac{C^K_{\phi\phi}(q=0,\omega=0)}{2{\rm Im}\left[C^R_{\phi\phi}(q=0,\omega\rightarrow 0)\right]}
 = \frac{2\bar{T}}{\omega}
\end{eqnarray}
and therefore assumed to be ${\bar T}>\omega$ is,
\begin{eqnarray}
\bar{T} = \frac{K_{neq}-2}{2K_{eq}}\label{T1}
\end{eqnarray}
where the energy-scales are expressed in units of $u/\alpha$, and length scales in units of $\alpha$.
The behavior of $C^K(q$=$0.5,\omega)$ and $2{\rm Im}\left[C^R_{\phi\phi}
(q=0.5,\omega)\right]$
are plotted in Fig.~\ref{fdt} for the equilibrium case when
$K_{eq}$=$K_{neq}$=$2$ and the non-equilibrium case of $K_{eq}$=$2,K_{neq}$=$3$.
Note that this ``temperature''
is dependent on the correlation function we use, contrary to the case of equilibrium
for which each ratio between response and fluctuation defines the same temperature.
Fig.~\ref{fdt} shows that besides the appearance of an effective temperature, a striking effect is
the appearance of a dissipation characterized by a non-zero slope of ${\rm Im}[C^R]
\propto -i\eta \omega$. As we shall show below, the temperature and the dissipation
already apparent at this stage will reappear in the RG analysis.

We now study how this non-equilibrium state is modified by a coupling between modes.
Although in principle any form of nonlinear coupling, such as for example a $\phi^4$ term
can be used, we focus here on the case of a $\cos(\gamma \phi)$ perturbation.
There are two reasons for such a choice:
i) If the phase $\phi$ represents real interacting bosons, the Hamiltonian cannot contain
perturbations coupled directly to $\phi$ but only to derivative or periodic functions
of $\phi$; ii) such a periodic term arises naturally when a periodic potential is added
on the system. It is the source of the Mott transition in one dimension~\cite{Giamarchibook},
and thus very natural to study in that context.
As for the case of equilibrium we study this term by a renormalization group (RG) procedure,
since the rest of the Keldysh action $S_0$ is quadratic. The Keldysh path-integral is
$Z_K = \int {\cal D}\left[\phi_{cl},\phi_q\right] e^{i \left(S_0 + S_{sg}\right)}$
where
\begin{equation}
S_{sg} = \frac{gu}{\alpha^2} \int dx \int dt\left[\cos{(\gamma \phi_-)}-\cos(\gamma \phi_+)\right]\label{Sg}
\end{equation}
Writing such an action assumes that after the quench has long taken place, one switches
on the cosine term infinitely slowly. Without the quench the system would thus relax to the ground
state (at $T$=$0$) \emph{in the presence} of the cosine term. To which state the system will tend
if the initial state is not in equilibrium is the very question we address here.
In order to perform an RG analysis, we split the modes between slow and fast components
$\phi_{cl,q}(xt) = \phi_{cl,q}^{<}(xt) + \phi_{cl,q}^{>}(xt)$
where
\begin{equation}
\begin{split}
\phi_{cl,q}^{<}(xt) &= \int_{-\infty}^{\infty}\frac{d\omega}{2\pi}\int_{-\Lambda^{\prime}}
^{\Lambda^{\prime}}\frac{dq}{2\pi} e^{i q x - i \omega t} \phi_{q,cl}(q,\omega) \\
\phi_{cl,q}^{>}(xt) &= \int_{-\infty}^{\infty}\frac{d\omega}{2\pi}\int_{\Lambda > |q| > \Lambda^{\prime}}
\frac{dq}{2\pi} e^{i q x - i \omega t} \phi_{q,cl}(q,\omega)
\end{split}
\end{equation}
and $\Lambda/\Lambda^{\prime} = e^{d(\ln l)}$, and integrate out the fast modes.
Care has to be taken to regularize the two-point functions appropriately, and
no cutoff on time should be imposed.
We choose here a standard momentum cutoff~\cite{Giamarchibook},
the details of the computation will be given elsewhere.

The first step of RG generates a correction to $S_{sg}$ as well as corrections to
$S_0$ which may be absorbed into
a redefinition of
$K$ and $u$. However when $K\neq K_0$, additional terms of the form
\begin{multline}
\delta S = \int dR \int d(ut) \frac{1}{\pi K}\left[- 2\frac{\eta}{u} \left(\frac{\Lambda}{\Lambda^{\prime}}\right)
\phi_q\partial_{ut} \phi_{cl}
\right. \\
\left. + i\frac{4 T_{eff} \eta}{u^2}\frac{K_0}{2K}\left(1 + \frac{K^2}{K_0^2}\right)
\left(\frac{\Lambda}{\Lambda^{\prime}}\right)^2 \phi_q^2\right]
\end{multline}
are generated.
\begin{figure}
\includegraphics[totalheight=5cm
]{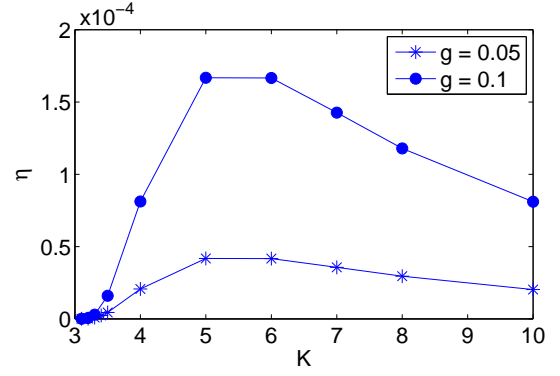}
\caption{\label{diss} Strength of the dissipation $\eta$ for $K_0$=$3$, $\gamma$=$2$ and $g$=$0.05$ and
$g$=$0.1$.}
\end{figure}
These corrections can be summarized in the following RG equations,
\begin{align}
\frac{d g}{d\ln l} &= \left[2 - \frac{\gamma^2}{8}K_0(1 + K^2/K_0^2)\right]g \label{gren}\\
\frac{d K^{-1}}{d \ln l} &= \!\! \frac{\pi g^2}{4\alpha^4}\left(\frac{\gamma^2}{2}\right)^2\!
\frac{K_0}{2}\left(1+\frac{K^2}{K_0^2}\right)I_K\label{Kren} \\
\frac{1}{Ku}\frac{du}{d\ln l} &= \!\! \frac{\pi g^2}{4\alpha^4}\left(\frac{\gamma^2}{2}\right)^2\!\!
\frac{K_0}{2}\left(1+\frac{K^2}{K_0^2}\right)I_u\label{uren} \\
\frac{d\eta}{d\ln l} &= \!\!  \eta + \frac{\pi g^2uK}{2\alpha^4}
\left(\frac{\gamma^2}{2}\right)^2
\!
\frac{K_0}{2}\left(1+\frac{K^2}{K_0^2}\right)I_{\eta}\label{etaren} \\
\frac{d(T_{eff}\eta)}{d\ln l} &= 2 T_{eff}\eta+\frac{\pi g^2u^2K^2}{4\alpha^4}\left(\frac{\gamma^2}{2}\right)^2
I_T \label{Tren}
\end{align}
where
$I_T$=$\int_{-\infty}^{\infty} dr\int_{-\infty}^{\infty} dt {\rm Re}
\left[ e^{-\frac{\gamma^2}{2}\langle (\phi_-(t,r) -\phi_+(0,0))^2\rangle}
F\right]$,
$I_{\eta}$=$\int_{-\infty}^{\infty} dr\int_{-\infty}^{\infty} dt \,t \,{\rm Im}
\left[ e^{-\frac{\gamma^2}{2}\langle (\phi_-(t,r) -\phi_+(0,0))^2\rangle}
F\right]$,
$I_{K/u}$=$\int_{-\infty}^{\infty} dr\int_{0}^{\infty} dt (r^2\mp t^2){\rm Im}
\left[ e^{-\frac{\gamma^2}{2}\langle (\phi_-(t,r) -\phi_+(0,0))^2\rangle}
F\right]$
and $F$=$\frac{1}{2}\left(e^{i\Lambda(t+r)}+ e^{i\Lambda(t-r)}\right)
+\frac{i}{2}\left(\frac{K_{eq}}{K_{neq}}-1\right) \left\{\sin(\Lambda(t+r))
+ \sin(\Lambda(t-r))\right\}$

Eq.~(\ref{gren}) reflects the new scaling dimension of the operator $e^{i\gamma \phi}$ due to the change
in the decay exponent from $K_{eq}$  to $K_{neq}$. It still defines two regimes, one in which
the cosine is
irrelevant, and one for which the perturbative RG would lead to strong coupling for the cosine term.
In equilibrium this would reflect the Berezinski-Kosterlitz-Thouless transition corresponding to
the Mott transition
($K=2$ and infinitesimal $g$).
Eq.~(\ref{Kren}) is the usual scaling of $K$, which is reduced by the presence of
the cosine term. This equation is also slightly
modified compared to the equilibrium one when $K_{eq}\neq K_{neq}$. Eq.~(\ref{uren})
is a renormalization of the velocity. It appears
here because we took a pure momentum cutoff which thus does not respect the Lorentz invariance.
It would appear also in equilibrium
with the same cutoff structure. These three equations would thus lead to two separate phases,
one in which the cosine is irrelevant,
and a strong coupling regime whose physics would be beyond the reach of the perturbative RG.
In order to stay in the regime for which the RG is reliable even asymptotically we
concentrate here on the case $K_{eq},K_{neq}>2$ for which the cosine
term is irrelevant according to Eq.~(\ref{gren}).
Other regimes of the phase diagram will be discussed elsewhere.
In this regime one could naively expect to recover the same physics as without
the cosine (namely the athermal state corresponding to Eq.~(\ref{S0a})).
However the two remaining equations
(\ref{etaren},\ref{Tren}) introduce qualitatively new physics and lead to quite a different state.

Eq.~(\ref{etaren}) shows that contrary to the case of an equilibrium
quantum system, for which the friction coefficient remains always infinitesimal ($\eta = 0^+$),
even at finite temperature, here
because of the combination of the cosine term and the initial out of equilibrium action,
a \emph{finite} friction is generated.
If one starts from the equilibrium situation $K$=$K_0$ then of course $I_{T,\eta}$=$0$
and one recovers the conventional results. The finite friction causes a crossover
of the mode dispersion at low energies
from a pure quantum behavior,
dominated by $(\partial_t\phi)^2 \to \omega^2 \phi^2$, to a more classical
one $\eta \partial_t \phi \to i \eta \omega \phi$, and the correlation functions will
reflect this.
Interestingly the physics of dissipation can also be recovered in a quench involving
fermions~\cite{Lancaster11}. Similar to the case studied here, an initial quench on a
system of noninteracting fermions can cause it to
reach a non-equilibrium steady state characterized by a highly broadened distribution function.
As a result switching on infinitesimal interactions (which can be treated within
the random phase approximation) can cause efficient scattering and
an enhanced particle-hole continuum which leads to a damping of collective modes,
at least for attractive interactions between fermions.

In addition to the generation of the friction,
Eq.~(\ref{Tren}) shows that a constant term (in the limit $\omega \to 0$)
is added to the Keldysh part of the action.
In this limit and in equilibrium, this term would simply be $\propto \omega \eta \coth \frac{\omega}{2T}
\xrightarrow{\omega=0} 2T \eta $.
Thus the constant term together with the appearance of
a \emph{finite} friction can be
interpreted as a finite temperature, at least for small enough frequencies.
One thus recovers at small frequencies the action (so called
Martin-Siggia-Rose action) of a classical system
with a finite friction and a thermal noise.
Note that because we have shown that the full action renormalizes to a \emph{quadratic} one, 
and that the Keldysh term tends to a constant, the temperature as defined above is indeed the one 
that will appear in \emph{all} $\phi$ correlations, at least asymptotically for low frequencies, 
contrary to the case of (\ref{S0a}). Therefore the non-linear coupling of the modes leads to 
a thermalization
of the system. The full frequency dependence of such a noise is however quite complicated
leading to an interesting crossover, depending on the frequency scale,
between the athermal distribution and the classical, finite temperature one at low frequencies.
In particular the RG flow itself has been derived from the quantum athermal correlations.
The corrections generated by the RG will thus change significantly
at a scale for which $\omega^2$=$\eta(\omega) \omega$. Since at this scale the system enters
in a more classical
regime with exponentially decaying correlation functions (see below),
this regime will not change the fact that
the cosine is irrelevant, and will simply slightly modify the final values of
the friction and temperature.

Fig.~\ref{diss} shows the solution for the renormalized
$\eta$ for two different $g$ and $K_0$=$3$. The non-monotonic
behavior arises because the larger is $K$ the more rapidly $g$ renormalizes to zero leading to
a smaller renormalized $\eta$. While
at the other end, when $K$=$K_0$, $\eta$=$0$. These two behaviors have to go through a maxima.
Quite naturally the friction is proportional to $g^2$. This is however not the case for the temperature
$T_{eff}$ which for small $g$ is found to reach the following value independent of $g$
(where we have set $F$=$1$ in the RG equations)
\begin{eqnarray}
T_{eff}^* = \frac{K_{neq}^*-2}{2K_{neq}^*}
\end{eqnarray}
While as $K_{neq}\rightarrow K_{eq}$, $\eta, \eta T_{eff} \rightarrow 0$,
$T_{eff}$ is non-zero.
This is because the order of limits
$\omega\rightarrow 0, K_{eq}\rightarrow K_{neq}$ do not commute.
Further $T_{eff}^*K_{neq}^*/K_{eq}^* \simeq {\bar T}$, and hence is
consistent with the non-interacting estimate for the temperature (\ref{T1}).

Let us finally compute the correlations at the thermal fixed point where the
action is (dropping $\omega^2$ terms in comparison to
$\omega\eta$)
\begin{eqnarray}
&&S^* = \sum_{q,\omega}\begin{pmatrix} \phi_{cl}^* & \phi_q^*\end{pmatrix}
\frac{1}{\pi K^*u} \nonumber \\
&&\!\!\begin{pmatrix} 0&& -i\eta^*\omega - u^2 q^2       \\
i \eta^*\omega - u^2 q^2&&
4iT_{eff}^*\eta^*\frac{K_0}{2K^*}\left(1 + \frac{K^{*2}}{K_0^2}\right)
\end{pmatrix}
\!\!\!\begin{pmatrix}
\phi_{cl}\\
\phi_q
\end{pmatrix}
\end{eqnarray}
The above implies that equal-time two point correlation functions decay exponentially in position,
\begin{equation}
\langle e^{i\phi_{cl}(x)}e^{-i\phi_{cl}(y)}
\rangle\simeq e^{-\frac{K_{neq}^*}{K_{eq}^*}T^*_{eff}\frac{\pi K^*}{u}|x-y|}
\end{equation}
while the dissipation affects unequal-time correlation functions
$G^R(q,t)$=$-\theta(t)(\pi K^* u/\eta^*) e^{-u^2q^2 t/\eta^*}$,
$G^K(q,t) = -(\frac{2 \pi i K^*}{uq^2})(\frac{T_{eff}^*K_{neq}^*}{K_{eq}^*}) e^{-u^2q^2 |t|/\eta^*}$.
Thus in an experiment involving a one dimensional Bose gas in a periodic potential~\cite{Bloch08}, 
a probe of the density-density response function which directly correspond to the correlators 
$\langle e^{i \gamma \phi} e^{-i \gamma \phi}\rangle$, should reveal 
the dissipative and thermal effects we predict.

In summary, by studying the particular example of a quenched Luttinger liquid in the
presence of a lattice, we have found a mechanism, that we believe is generic by which a
non-equilibrium system in the
presence of mode coupling will both thermalize
and acquire a finite friction or lifetime for the modes.
It is important to note that these effects are related to the presence of a continuum
of excitations in the system by which local degrees of freedom can relax and exchange energy.
By this argument it is possible that thermalization might not occur in
the Mott insulator phase, and the fact that
$T_{eff}$ vanishes near the critical point, might
be a prelude to this physics. An investigation of this issue and also how
the results depend upon the rapidity with which the cosine potential is switched on, 
are important open questions left for future studies.

{\sl Acknowledgements}:
This work was supported by NSF-DMR (Grant No. 1004589)
and by the Swiss SNF under MaNEP and Division II. We
thank E. Altman and E. Dalla Torre for useful discussions.


\begin{thebibliography}{15}
\expandafter\ifx\csname natexlab\endcsname\relax\def\natexlab#1{#1}\fi
\expandafter\ifx\csname bibnamefont\endcsname\relax
  \def\bibnamefont#1{#1}\fi
\expandafter\ifx\csname bibfnamefont\endcsname\relax
  \def\bibfnamefont#1{#1}\fi
\expandafter\ifx\csname citenamefont\endcsname\relax
  \def\citenamefont#1{#1}\fi
\expandafter\ifx\csname url\endcsname\relax
  \def\url#1{\texttt{#1}}\fi
\expandafter\ifx\csname urlprefix\endcsname\relax\def\urlprefix{URL }\fi
\providecommand{\bibinfo}[2]{#2}
\providecommand{\eprint}[2][]{\url{#2}}

\bibitem[{\citenamefont{Bloch et~al.}(2008)\citenamefont{Bloch, Dalibard, and
  Zwerger}}]{Bloch08}
\bibinfo{author}{\bibfnamefont{I.}~\bibnamefont{Bloch}},
  \bibinfo{author}{\bibfnamefont{J.}~\bibnamefont{Dalibard}}, \bibnamefont{and}
  \bibinfo{author}{\bibfnamefont{W.}~\bibnamefont{Zwerger}},
  \bibinfo{journal}{Rev. Mod. Phys.} \textbf{\bibinfo{volume}{80}},
  \bibinfo{pages}{885} (\bibinfo{year}{2008}).

\bibitem[{\citenamefont{Dalla~Torre et~al.}(2010)\citenamefont{Dalla~Torre,
  Demler, Giamarchi, and Altman}}]{Torre10}
\bibinfo{author}{\bibfnamefont{E.~G.} \bibnamefont{Dalla~Torre}},
  \bibinfo{author}{\bibfnamefont{E.}~\bibnamefont{Demler}},
  \bibinfo{author}{\bibfnamefont{T.}~\bibnamefont{Giamarchi}},
  \bibnamefont{and} \bibinfo{author}{\bibfnamefont{E.}~\bibnamefont{Altman}},
  \bibinfo{journal}{Nat. Phys} \textbf{\bibinfo{volume}{6}},
  \bibinfo{pages}{806} (\bibinfo{year}{2010}).

\bibitem{Polkovrev09} A. Polkovnikov {\it et al},
Rev. Mod. Phys. {\bf 83}, 863 (2011). 


\bibitem[{\citenamefont{Rigol et~al.}(2007)\citenamefont{Rigol, Dunjko,
  Yurovsky, and Olshanii}}]{Rigol07}
\bibinfo{author}{\bibfnamefont{M.}~\bibnamefont{Rigol}},
  \bibinfo{author}{\bibfnamefont{V.}~\bibnamefont{Dunjko}},
  \bibinfo{author}{\bibfnamefont{V.}~\bibnamefont{Yurovsky}}, \bibnamefont{and}
  \bibinfo{author}{\bibfnamefont{M.}~\bibnamefont{Olshanii}},
  \bibinfo{journal}{Phys. Rev. Lett.} \textbf{\bibinfo{volume}{98}},
  \bibinfo{pages}{050405} (\bibinfo{year}{2007}).

\bibitem[{\citenamefont{Cazalilla}(2006)}]{Cazalilla06}
A. Iucci and M. A. Cazalilla,
Phys. Rev. A {\bf 80}, 063619 (2009).

\bibitem[{\citenamefont{Calabrese and Cardy}(2006)}]{Calabrese06}
\bibinfo{author}{\bibfnamefont{P.}~\bibnamefont{Calabrese}} \bibnamefont{and}
  \bibinfo{author}{\bibfnamefont{J.}~\bibnamefont{Cardy}},
  \bibinfo{journal}{Phys. Rev. Lett.} \textbf{\bibinfo{volume}{96}},
  \bibinfo{pages}{136801} (\bibinfo{year}{2006}).

\bibitem[{\citenamefont{Kollath et~al.}(2007)\citenamefont{Kollath, L\"auchli,
  and Altman}}]{Kollath07}
\bibinfo{author}{\bibfnamefont{C.}~\bibnamefont{Kollath}},
  \bibinfo{author}{\bibfnamefont{A.~M.} \bibnamefont{L\"auchli}},
  \bibnamefont{and} \bibinfo{author}{\bibfnamefont{E.}~\bibnamefont{Altman}},
  \bibinfo{journal}{Phys. Rev. Lett.} \textbf{\bibinfo{volume}{98}},
  \bibinfo{pages}{180601} (\bibinfo{year}{2007}).

\bibitem[{\citenamefont{Barthel and Schollw\"ock}(2008)}]{Barthel08}
\bibinfo{author}{\bibfnamefont{T.}~\bibnamefont{Barthel}} \bibnamefont{and}
  \bibinfo{author}{\bibfnamefont{U.}~\bibnamefont{Schollw\"ock}},
  \bibinfo{journal}{Phys. Rev. Lett.} \textbf{\bibinfo{volume}{100}},
  \bibinfo{pages}{100601} (\bibinfo{year}{2008}).

\bibitem[{\citenamefont{Gangardt and Pustilnik}(2008)}]{Gangardt08}
\bibinfo{author}{\bibfnamefont{D.~M.} \bibnamefont{Gangardt}} \bibnamefont{and}
  \bibinfo{author}{\bibfnamefont{M.}~\bibnamefont{Pustilnik}},
  \bibinfo{journal}{Phys. Rev. A} \textbf{\bibinfo{volume}{77}},
  \bibinfo{pages}{041604} (\bibinfo{year}{2008}).

\bibitem[{\citenamefont{Rossini et~al.}(2010)\citenamefont{Rossini, Suzuki,
  Mussardo, Santoro, and Silva}}]{Rossini10}
\bibinfo{author}{\bibfnamefont{D.}~\bibnamefont{Rossini}},
  \bibinfo{author}{\bibfnamefont{S.}~\bibnamefont{Suzuki}},
  \bibinfo{author}{\bibfnamefont{G.}~\bibnamefont{Mussardo}},
  \bibinfo{author}{\bibfnamefont{G.~E.} \bibnamefont{Santoro}},
  \bibnamefont{and} \bibinfo{author}{\bibfnamefont{A.}~\bibnamefont{Silva}},
  \bibinfo{journal}{Phys. Rev. B} \textbf{\bibinfo{volume}{82}},
  \bibinfo{pages}{144302} (\bibinfo{year}{2010}).

\bibitem[{\citenamefont{Roux}(2009)}]{Roux09}
\bibinfo{author}{\bibfnamefont{G.}~\bibnamefont{Roux}}, \bibinfo{journal}{Phys.
  Rev. A} \textbf{\bibinfo{volume}{79}}, \bibinfo{pages}{021608}
  (\bibinfo{year}{2009}).

\bibitem[{\citenamefont{Biroli et~al.}(2010)\citenamefont{Biroli, Kollath, and
  L\"auchli}}]{Biroli10}
\bibinfo{author}{\bibfnamefont{G.}~\bibnamefont{Biroli}},
  \bibinfo{author}{\bibfnamefont{C.}~\bibnamefont{Kollath}}, \bibnamefont{and}
  \bibinfo{author}{\bibfnamefont{A.~M.} \bibnamefont{L\"auchli}},
  \bibinfo{journal}{Phys. Rev. Lett.} \textbf{\bibinfo{volume}{105}},
  \bibinfo{pages}{250401} (\bibinfo{year}{2010}).

\bibitem[{\citenamefont{Giamarchi}(2004)}]{Giamarchibook}
\bibinfo{author}{\bibfnamefont{T.}~\bibnamefont{Giamarchi}},
  \bibinfo{journal}{{\sl Quantum Physics in One Dimension}, Oxford University
  Press, Oxford}  (\bibinfo{year}{2004}).

\bibitem[{\citenamefont{Kamenev}(2005)}]{Kamenevrev}
\bibinfo{author}{\bibfnamefont{A.}~\bibnamefont{Kamenev}},
  \bibinfo{journal}{Nanophysics: Coherence and Transport, Les Houches 2004
  session No. LXXX1 (Elsevier, Amsterdam)}  (\bibinfo{year}{2005}).

\bibitem{Lancaster11} J. Lancaster, T. Giamarchi and A. Mitra, Phys. Rev. B {\bf 84}, 075143
(2011).



\end{thebibliography}

\end{document}